\documentclass[twocolumn]{revtex4-2}
\usepackage[dvipdfmx]{graphicx}
\usepackage{amsmath,amsbsy,amssymb}
\usepackage{bm}
\usepackage{mathrsfs}
\usepackage{ulem}
\usepackage{textgreek}
\usepackage{mathrsfs}
\usepackage{multirow}
\usepackage{physics}
\usepackage{comment}

\usepackage{color}

\newcommand{\diff}{\mathrm{d}}

\newcommand{\imu}{\mathrm{i}}

\newcommand{\ua}{\uparrow}
\newcommand{\da}{\downarrow}
\newcommand{\dg}{\dagger}
\newcommand{\la}{\langle}
\newcommand{\ra}{\rangle}
\newcommand{\al}{\alpha}
\newcommand{\sg}{\sigma}
\newcommand{\gm}{\gamma}

\newcommand{\dvec}[1]{\hspace{-1mm}\stackrel{\leftrightarrow}{#1}\hspace{-1mm}}

\begin{document}

\title{
Spin current and chirality degrees of freedom inherent in localized electron orbitals
}

\author{
Shintaro Hoshino$^1$, Michi-To Suzuki$^{2,3}$, and Hiroaki Ikeda$^4$
}

\affiliation{
$^1$Department of Physics, Saitama University, Sakura, Saitama 338-8570, Japan 
\\
$^2$Center for Computational Materials Science, Institute for Materials Research, Tohoku University, Sendai, Miyagi 980-8577, Japan
\\
$^3$Center for Spintronics Research Network, Graduate School of Engineering Science,
Osaka University, Toyonaka, Osaka 560-8531, Japan
\\
$^4$Department of Physics, Ritsumeikan University, Kusatsu, Shiga 525-8577, Japan
}

\date{\today}

\begin{abstract}
In solid state physics, any phase transition is commonly observed as a change in the microscopic distribution of charge, spin, or current. Here we report the nature of an exotic order parameter inherent in the localized electron orbitals that cannot be primarily captured by these three fundamental quantities. 
This order parameter is described as the electric toroidal multipoles connecting different total angular momenta under the spin-orbit coupling. The corresponding microscopic physical quantity is the spin current tensor on an atomic scale, which induces spin-derived electric polarization and the chirality of the Dirac equation.
We stress that the chirality intrinsic to the elementary particle is the essence of electric toroidal multipoles.
These findings link microscopic spin currents and chirality in the Dirac theory to the concept of multipoles and provide a new perspective for quantum states of matter.
\end{abstract}

\maketitle

{\it Introduction.---}
In strongly correlated electron systems, the spin and orbital degrees of freedom of nearly localized electrons are activated by electronic correlations, resulting in various intriguing phenomena, such as heavy electrons, unconventional superconductivity, and exotic magnetic orders.
These quantum states are commonly characterized by the 
spatial distribution of the fundamental microscopic physical quantities, charge,  spin, and current.
They have been systematically studied using the concept of multipole expansions
\cite{Ohkawa83,Shiina97,Kuramoto00,Santini00,Kiss05,Takimoto05,Kusunose08,Kuramoto09,Haule09,Ikeda12,Suzuki17,Hayami18}.
These multipoles are classified into four categories (electric, magnetic, magnetic toroidal, and electric toroidal) according to spatial parity and time-reversal parity \cite{Hayami18}. 
Such a classification is useful for elucidating exotic orders and predicting novel response to external fields.

In particular, electric toroidal multipole ordering has recently attracted much attention as novel nonmagnetic degree of freedom \cite{Hayami19,Hirose22,Hayami21,Oiwa22,Kishine22} 
which
has not been strongly recognized so far.
In localized electron systems, such degrees of freedom are inherent in the components connecting different total angular momenta under the spin-orbit coupling $\lambda \bm \ell \cdot \bm s$. 
The simplest rank-$1$ electric toroidal dipole is 
written as $\bm G(1) = \bm \ell \times \bm s$ \cite{Wang17,Kusunose20,Chikano21,Hayami21}. 
When considering this electric toroidal dipole in analogy to a magnetic toroidal dipole \cite{Spaldin08}, we would expect the electric polarization to be circularly aligned.
The ordinary electric polarization $\bm P(\bm r)=\bm r\rho(\bm r) $ ($\rho$ is a charge density), however, cannot capture the toroidal structure,
as $\bm P(\bm r)$ is a simple charge distribution.
What physical quantity characterizes the nature of electric toroidal multipoles microscopically?

In the following,
we clarify the underlying fundamental physical quantity using the knowledge of the relativistic quantum mechanics.
It is shown that the electric polarization $\bm P_S(\bm r)$ induced by the spin degrees of freedom is responsible for the electric toroidal moment.
The spin-derived electric polarization $\bm P_S(\bm r)$ is also associated with the microscopic spin current tensor, which is regarded as one of the fundamental physical quantities analogous to charge, spin, and electric current.
The relation between electric polarization and spin current has been discussed in the context of spintronics based on weak-coupling itinerant Fermi liquid description~\cite{Wang06}.
We further find the relation between the electric toroidal multipole, spin current, and chirality of the Dirac equation by employing the localized electron picture.
This chirality degree of freedom is intimately related to the diagonal part of the spin current, and is a more fundamental quantity corresponding to electric toroidal multipoles.
We thus link microscopic spin currents and chirality in Dirac theory to the concept of multipoles.

\begin{table*}[t]
  \begin{tabular}{ccccccccccc}
\hline
& 
& 
& $\rho$ & $\bm M_S$ & $\bm j$ & $\bm P_S$ & $\tau^Z$ & $\tau^X$ ($\sim \rho$) 
&\ \ \ $\tau^Y\ (\sim \bm\nabla\cdot \bm M_S)$ \ \ \ 
&$\bm \nabla \cdot \bm P_S$
\\ 
Multipole &Type & SI/TR 
& $+/+$ & $+/-$ & $-/-$ & $-/+$ & $-/+$ & $+/+$&$-/-$&$+/+$
\\ 
\hline
Electric Toroidal & $(2q+1)_d$  & $+/+$ & 0 & 0 & 0& Nonzero &Nonzero&0&0&0\\
Magnetic Toroidal & $(2q)_d$  & $+/-$ & 0 & Nonzero & 0 & 0 &0&0&0&0\\
 \hline
Electric &$(2q)_{a,b,c}$  & $+/+$  & Nonzero & 0 & 0& Nonzero &0&Nonzero&0&Nonzero\\
Magnetic &$(2q+1)_{a,b,c}$ & $+/-$ & 0 & Nonzero & Nonzero& 0 &0&0&Nonzero&0\\
\hline
  \end{tabular}
 \caption{
List of the multipole distribution function
$f(\bm r; p_\eta, \gm)$ ($p=2q$ or $p=2q+1$)
defined in Eq.~\eqref{eq:def_multipole_basis} where 
$f=\rho,M_{S\mu},j_\mu,P_{S\mu},\tau^{Z,X,Y}$, and $\bm \nabla \cdot \bm P_S$.
The label `Nonzero' (`0') for each cell indicates a non-zero (zero) multipole distribution function in the leading-order contribution of the non-relativistic limit.
The signs of the spatial inversion (SI) $\mathcal P$ and time-reversal (TR)
$\Theta$ indicated in the third column and the second row are defined as $\mathcal P f(\bm r)\mathcal P^{-1} = \pm f(-\bm r)$ and $\Theta f(\bm r)\Theta^{-1} = \pm f(\bm r)$.
}
\label{tab:classification}
\end{table*}

{\it Definition of multipoles.---}
Let us start with the definition of multipoles in localized electron orbitals with the angular momentum $\ell$.
The multipole operators are defined as a complete matrix basis set to describe all operators of the type $c_{m\sg}^\dg c_{m'\sg'}$, where $c_{m\sg}$ is the annihilation operator of the electron with the magnetic quantum number $m\in[-\ell,\ell]$ and spin $\sg =\ua,\da$.
Under strong spin-orbit coupling, the multipole operators are usually considered only for a ground-state $j=\ell\pm 1/2$ multiplet.
We here consider the full space containing different $j$ multiplets.
The classification scheme recently formulated for $\ell=1$ case \cite{Chikano21} is applied to the general $\ell$ \cite{suppl}.
In this case, the generic rank-$p$ multipole is written as
\begin{align}
X^\gm(p_\eta) &= \sum_{mm'\sg\sg'} c^\dg_{m\sg} \mathcal O^\gm_{m\sg,m'\sg'}(p_\eta) c_{m'\sg'}
,
\end{align}
where $p=0,\cdots,2\ell+1$, and $\mathcal O$ is a matrix representation of the multipole. 
$\gm$ is a label to distinguish $2p+1$ degeneracies.
We have separated the rank-$p$ multipole into several pieces denoted by the $\eta$-index:
$\eta=a,b$ denote intra-$j$ multiplet component, and $\eta=c,d$ corresponds to a component connecting different $j$ multiplet \cite{matrixO}. 
For $\eta=a,b,c$, 
the even ($p=2q$) or odd ($p=2q+1$) rank coincides with the even/odd of the time reversal (TR), but this relation is reversed for $d$ components.
Note that in the present  intra-$\ell$ case, the parity of spatial inversion (SI) is always even.
In addition, the type $\eta=d$ multipoles are categorized as the toroidal multipoles from its symmetry \cite{Kusunose20}.
These results are summarized in the left three columns (`Multipole', `Type', `SI/TR') of Tab.~\ref{tab:classification}.

{\it Charge and current distribution.---}
Since the multipoles obtained above were defined based on localized electron orbitals, they can be described as microscopic charge or current distributions in continuous space.
With the field operator $\psi_\sg(\bm r)$,
 the microscopic charge density is given by $ \rho_0 = e \psi^\dg \psi$, and the current density and magnetization operators by
\begin{align}
    \bm j &= \frac{e}{2m} \psi^\dg  \dvec{\bm p} \psi , \ \ \
    \bm M_S = \frac{\hbar e}{2mc} \psi^\dg \bm \sg \psi
    , \label{eq:j_MS_def}
\end{align}
where $\bm p= -\imu \hbar \bm\nabla$, and $A\dvec{\partial} B = A\partial B - (\partial A)B$.
The spin summation is implicitly performed.
We expand the operator by the atomic orbitals as $\psi_\sg(\bm r) = \sum_{m} R(r) Y_{\ell m}(\hat{\bm r}) c_{m\sg}$. 
Then, for example, the current operator is written as 
\begin{align}
    \bm j(\bm r) = \sum_{p\eta \gm} X^\gm(p_\eta) \bm j(\bm r ; p_\eta, \gm)
    . \label{eq:def_multipole_basis}
\end{align}
Namely, once the multipoles are given, the spatial distribution of the current can be visualized through the product with the distribution function, $\bm j(\bm r;p_\eta,\gm)$ for each set of $(p_\eta,\gm)$.
Similar expansions are possible for other microscopic physical quantities.

However, as will be shown later, there is no primary change in the spacial distributions of $\rho_0(\bm r)$, $\bm j(\bm r)$, and $\bm M_S(\bm r)$ in the ordered state of electric toroidal dipole $\bm G(1)$. 
In order to obtain another microscopic quantity having primary change, we start with the basic Hamiltonian $\mathscr H$ with relativistic corrections. 
Then the current and charge are obtained by $\bm j_{\rm tot} = - \frac{\delta \mathscr H}{\delta \bm A/c}$ and $\rho_{\rm tot} = \frac{\delta \mathscr H}{\delta \Phi}$, where $\bm A$ and $\Phi$ are vector and scalar potentials \cite{Wang06,suppl}.
Based on the spin dependence, the total current and charge can be uniquely separated
into two parts:
$\bm j_{\rm tot}= \bm j + c \bm \nabla \times \bm M_S$ and $\rho_{\rm tot} = \rho - \bm \nabla \cdot \bm P_S$.
The current density and magnetization have already been defined above.
The charge density and electric polarization are given by
\begin{align}
\rho &=\qty( 1+\frac{\hbar^2 }{8m^2c^2}\bm \nabla^2) e\psi^\dg\psi \simeq e\psi^\dg\psi = \rho_0 
,\\
    \bm P_S &= \frac{\hbar e}{8m^2c^2} \psi^\dg  \dvec{\bm p} \times \bm \sg \psi 
 .   \label{eq:p_def}
\end{align}
The second term in $\rho$ originates from the uncertainty of the position in relativistic quantum mechanics.
However, this second term has only a minor correction on the charge distribution since it originates from the second derivative of the large first term ($\rho_0$), and the spin degrees of freedom are not directly involved.
This correction term has the same origin as the Darwin term in the Hamiltonian \cite{Baym_book}, which only affects the $s$ electron ($\ell=0$) and vanishes for $\ell>0$.

Here we make a few comments on the spin-derived electric polarization $\bm P_S$.
The Gordon decomposition of the charge and current also introduces the microscopic magnetization and electric polarization \cite{Sakurai_book,Baym_book}.
We also note that the presence of the microscopic electric polarization has the same physical origin as the Aharonov-Casher effect \cite{Aharonov84}, in which the particles with magnetization are affected by the electric field.

The present spin-derived electric polarization $\bm P_S$ is exactly related to microscopic spin current.
The connection between the spin current and electric polarization has been discussed for the non-colinear magnets \cite{Katsura05} and the electron gas model in the context of the spintronics \cite{Wang06}.
The spin-derived electric polarization $\bm P_S$ can be further rewritten as
\begin{align}
P_{S\mu} = \frac{\hbar^2 e}{8m^2c^2} \epsilon_{\mu\nu\lambda} j_{S\nu\lambda},
\end{align}
where the spin current 
$\displaystyle j_{S\mu\nu} = -\imu \psi^\dg \dvec{\partial}_\mu \sg^\nu \psi$
and the antisymmetric tensor $\epsilon_{\mu\nu\lambda}$ are introduced ($\mu,\nu,\lambda=x,y,z$) \cite{Wang06}.
Furthermore, the spin current tensor $j_{S\mu\nu}$
may be classified into the components with rank $0$ (pseudoscalar), $1$, and $2$ \cite{Hayami18}.
We find that the rank-$0$ pseudoscalar component is related to the chirality degrees of freedom in the Dirac equation.
The chirality density in the second-quantized form is defined by the annihilation operators for right- and left-handed chiral fermions, $\psi_{R,L}$, in the Weyl basis as follows,
\begin{align}
\tau^Z = \psi^\dg_R\psi_R - \psi_L^\dg \psi_L ,
\label{eq:Dirac_chirality}
\end{align}
where the subscript $R$ and $L$ denote the chirality degrees of freedom.
The chirality density is evaluated in the non-relativistic limit as follows \cite{Berestetskii_book, suppl},
\begin{align}
    \tau^Z &\simeq \frac{1}{2mc}\psi^\dg \dvec{\bm p} \cdot \bm \sg\psi = \frac{\hbar}{2mc} j_{S\mu\mu} .
    \label{eq:rewrite_Dirac_chirality}
\end{align}
Namely, the pseudoscalar part of the spin current tensor represents the chirality density operator in the non-relativistic limit.
Equation ~\eqref{eq:rewrite_Dirac_chirality} has a structure like the helicity, i.e., how much the moving direction of the electron is aligned along the direction of its spinning axis.
In condensed matter physics, the chiral magnetic anisotropy effect is expected, if the chirality density integrated over the bulk has a finite expectation value \cite{Rikken01,Tokura18}.

The chirality operator is further supplemented by 
another
operators related to the chirality degrees of freedom $R,L$:
\begin{align}
    \tau^X& = \psi_R^\dg \psi_L + \psi_L^\dg \psi_R \simeq \frac{1}{e} \rho_0
    , \label{eq:Lorentz_scalar}
    \\
    \tau^Y &= - \imu (\psi_R^\dg \psi_L - \psi_L^\dg \psi_R) \simeq \frac{1}{e}\bm \nabla \cdot \bm M_S
    , \label{eq:Lorentz_pseudoscalar}
\end{align}
which correspond to the Lorentz scalar and pseudoscalar \cite{Berestetskii_book,suppl}, respectively, and
the right-most sides in Eqs. (\ref{eq:Lorentz_scalar}) and (\ref{eq:Lorentz_pseudoscalar}) express non-relativistic limits.
Equations~\eqref{eq:Dirac_chirality}, \eqref{eq:Lorentz_scalar}, \eqref{eq:Lorentz_pseudoscalar} are represented by as a pseudospin in the `chirality space'.
In the non-relativistic limit, the $R$ and $L$ components almost equally exist, leading to the dominant $\la \tau^X\ra$ component.

All of the multipole distribution functions now represent the microscopic physical quantities, i.e., $\rho$, $\bm M_S$, $\bm j$, $\bm P_S$, $\tau^{X,Y,Z}$. They are summarized in Tab.~\ref{tab:classification} 
with their signs from parity for spatial inversion and time-reversal transformations.
As shown in Tab.~\ref{tab:classification}, the electric toroidal moment $G(2q+1)\equiv X\big( (2q+1)_d \big)$ appears only with
the chirality density operator $\tau^Z$ and electric polarization $\bm P_S$.
It is also notable that the divergence $\bm \nabla \cdot \bm P_S$ is absent, which implies the rotating nature. 
In this context, it is interesting to compare it with the magnetic toroidal moment $T(2q) \equiv X\big((2q)_d \big)$, which is not accompanied by $\tau^Y \sim \bm \nabla \cdot \bm M_S$.

{\it Coupling to external fields.---}
Once the magnetization and electric polarization are given, the coupling to the external electromagnetic fields in the Hamiltonian 
is written as
\begin{align}
    \mathscr H_{\rm ext} 
    &= \int \diff \bm r\qty(
    \rho \Phi_{\rm ext}
     - \frac{1}{c} \bm j \cdot \bm A_{\rm ext}
     - \bm M_S \cdot \bm B_{\rm ext}
     - \bm P_S \cdot \bm E_{\rm ext}
    )
    \label{eq:emcoupling}
\end{align}
in the linear response.
Here the external magnetic field $\bm B_{\rm ext} = \bm \nabla \times \bm A_{\rm ext}$ and the electric field
$\bm E_{\rm ext} = -\bm \nabla \Phi_{\rm ext}$ are introduced.
In the atomic limit model with fixed $\ell$, the uniform electric field does not couple with the electric polarization directly since $\int \diff \bm r \bm P_S(\bm r) = 0$.
We need the electric field modulating on an atomic scale for the finite coupling to the electric polarization in the present setup.

The spatially uniform fields lead to the familiar form of the contribution to magnetization and electric polarization.
It is natural (but not unique) to define the potentials as $\Phi_{\rm ext}(\bm r) = - \bm E_0 \cdot \bm r$ and $\bm A_{\rm ext}(\bm r) =\frac 1 2 \bm B_0 \times \bm r$.
We then have another expression 
$    \mathscr H_{\rm ext} = - \int \diff \bm r\qty(
    \bm M \cdot \bm B_0 + \bm P \cdot \bm E_0
    )
$ 
where the magnetization and electric polarization are
    $\bm M=\frac{e}{4mc} \psi^\dg \dvec{\bm L} \psi +\bm M_S$
    and 
    $\bm P = \bm r\rho + \bm P_S
    $
with $\bm L = \bm r\times \bm p$.
This expression provides an intuitive understanding of magnetization and polarization.

{\it Analysis of the $\ell=1$ model.---}
To gain more insight, let us consider a mean-field Hamiltonian in the simple $\ell=1$ localized model, given by
$\mathscr H_{\rm MF} = \sum_i \left[ \lambda (\bm \ell \cdot \bm s)_i -\bm h_i \cdot \bm G_i \right]$ where $i$ is a site index.
$\lambda$ is the spin-orbit coupling, $\bm G_i \equiv (\bm \ell \times \bm s)_i = \sqrt 2 \, \bm X_i (1_d)$ is an electric toroidal multipole, and $\bm h_i = \sum_j J_{ij}\la \bm G_j \ra$ is the local mean-field generated by the spontaneous symmetry breaking with the exchange interaction $J_{ij}$ \cite{Iwazaki21}.

Hereafter we clarify the physical consequence of the electric toroidal moment. For simplicity, the site index $i$ is omitted.
Without loss of generality, we can choose $\bm h = h\hat {\bm z}$ ($h\geq 0$).
The ground-state wave function is written as
$
    |\psi_{\pm} \ra = \imu\al |\tfrac 3 2, \pm \tfrac 1 2 \ra
    + |\tfrac 1 2 , \pm \tfrac 1 2 \ra \, ,
$
where $\al = \frac{3\lambda - \sqrt{9\lambda^2 + 8h^2}}{2\sqrt 2 \, h} \simeq - \frac{2\sqrt 2\, h}{3\lambda}$ ($|\al| \ll 1$)
and the ground state energy is $E= - \frac 1 4 (\lambda + \sqrt{9\lambda^2 + 8h^2} )$.
The single electron state is written as $|jj_z\ra = \sum_{m\sg}\la 1m\frac 1 2 \sg | jj_z\ra c_{m\sg}^\dg |0\ra$.
The ground-state degeneracy is protected by the time-reversal symmetry.
In the ordered state, any physical quantities can be calculated with the density matrix $\hat \rho = |\psi_+ \ra \la \psi_+ | + |\psi_- \ra \la \psi_- |$.
First, we evaluate the electric charge and spin current as
\begin{align}
    \frac{\la \rho_0(\bm r) \ra}{e R^2(r)}
    &=  \frac{(2+\al^2)r^2 + 3\al^2z^2}{8\pi r^2(1+\al^2)}
    , \label{eq:charge_conc}
    \\
    \frac{\la \bm j_{Sx}(\bm r) \ra}{R^2(r)} &= - \frac{(1-\al^2)\bm C_x - \frac{3}{\sqrt 2}\al \bm C_y}{2\pi r(1+\al^2)}
    , \label{eq:jsx}
    \\
    \frac{\la \bm j_{Sy}(\bm r) \ra}{R^2(r)} &= - \frac{(1-\al^2)\bm C_y + \frac{3}{\sqrt 2}\al \bm C_x}{2\pi r(1+\al^2)}
    , \label{eq:jsy}
    \\
    \frac{\la \bm j_{Sz}(\bm r)\ra}{R^2(r)} &= - \frac{(1+\frac 1 2 \al^2)\bm C_z}{2\pi r(1+\al^2)}
    , \label{eq:jsz}
\end{align}
where $r^2=x^2+y^2+z^2$ and $R(r)$ is a radial wave function.
The three-component vector is defined by $(\bm j_{S\mu})_\nu = j_{S\nu\mu}$, which represents a current vector of the spin $S^\mu$.
We have defined the vector field 
$\bm C_z(\bm r) = (x\hat{\bm y} - y \hat{\bm x})/r$, 
which is circularly rotating around the $z$-axis.
The $x$ and $y$ components ($\bm C_{x,y}$) are also introduced similarly.
The magnetization $\bm M_S$ and the electric current $\bm j$ remains zero since the time-reversal symmetry is preserved.
We emphasize that the charge distribution has only $O(\al^2)$ contribution, and the leading-order $O(\al)$ contribution is absent as shown in Eq.~\eqref{eq:charge_conc}.
The primary order parameter cannot be captured by the change of the charge distribution.

\begin{figure}[t]
\begin{center}
\includegraphics[width=85mm]{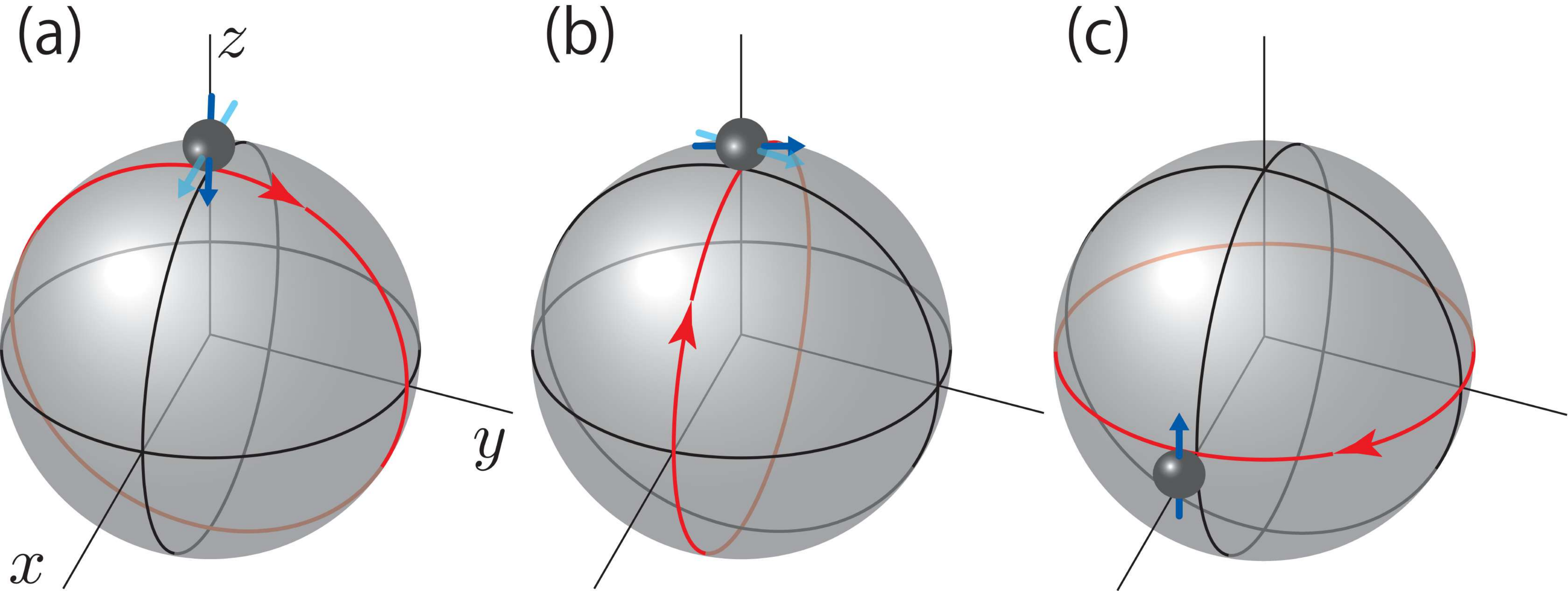}
\caption{
Schematic picture for the spin-current (a) $\la \bm j_{Sx}\ra$, (b) $\la \bm j_{Sy}\ra$, and (c) $\la \bm j_{Sz}\ra $ on a unit sphere, which are based on Eqs.~(\ref{eq:jsx}--\ref{eq:jsz}).
In the presence of the $\la G^z\ra $ order parameter, the additional torque tilts the spin direction as in (a) and (b).
}
\label{fig:spin_current}
\end{center}
\end{figure}

Figure~\ref{fig:spin_current} shows a schematic illustration of the spin current distribution given by Eqs.~\eqref{eq:jsx}--\eqref{eq:jsz}.
If there is no internal mean-field, i.e., $\al=0$, the spin current direction is perpendicular to the direction of the spin (see Fig.~\ref{fig:spin_current}), because the spin-orbit coupling minimizes the inner product $\bm \ell \cdot \bm s$.
On the other hand, for $\al\neq 0$, $\la \bm j_{Sx} \ra$ and $\la \bm j_{Sy}\ra$ components qualitatively change: the current and spin directions are not perpendicular to each other.
With $|\al|\ll 1$ in mind, we can rotate the spin in the $xy$-plane and redefine the current $\bm j_{Sx'}$ and $\bm j_{Sy'}$ [from the light-blue to deep-blue arrows in Fig.~\ref{fig:spin_current}(a,b)], so that these current directions remain unchanged with respect to the $\al=0$ case.
This behavior reflects the presence of $\bm \ell \times \bm s$.

\begin{figure}[t]
\begin{center}
\includegraphics[width=90mm]{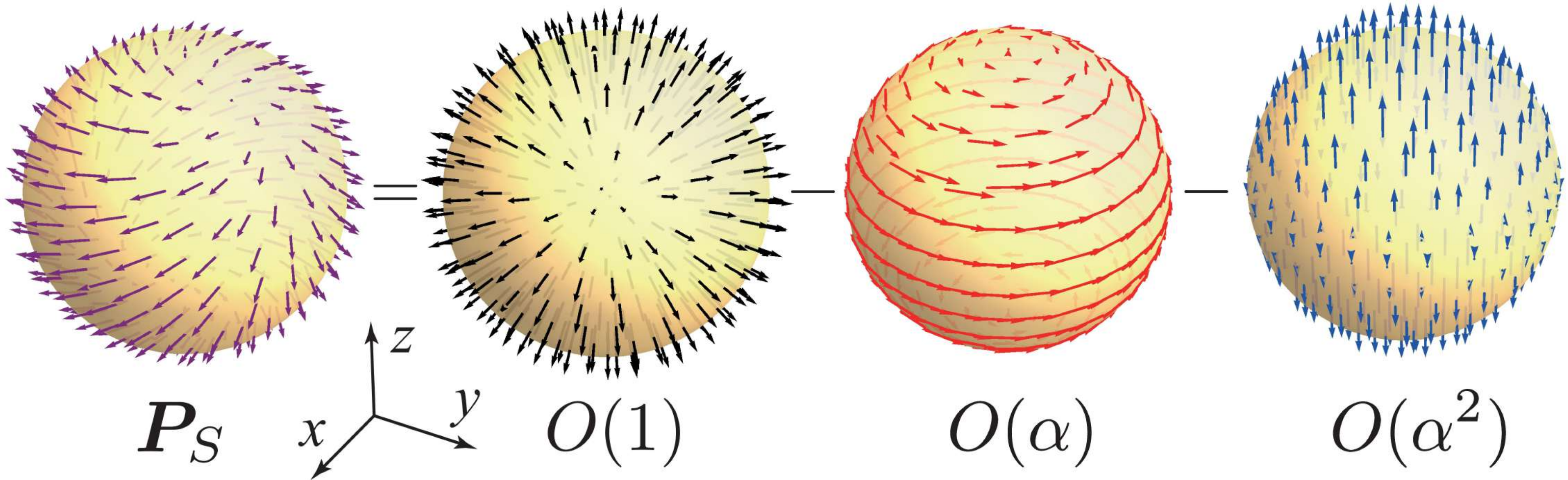}
\caption{
Schematic picture for the spatial distribution of the electric polarization $\bm P_S(\bm r)$ on a unit sphere.
In the right-hand side, the contributions from $O(1)$, $O(\al)$, $O(\al^2)$ [See Eq.~\eqref{eq:polarization_concrete}] in the $\al\to 0$ limit are separately plotted. 
}
\label{fig:polarization}
\end{center}
\end{figure}

We can also explicitly evaluate the expectation value of the electric polarization around the specified atomic site as
\begin{align}
    \frac{\la \bm P_S(\bm r)\ra}{R^2(r)} &= \frac{\hbar^2 e}{8m^2c^2} \, \frac{(-2+\frac{\al^2}{2})\hat {\bm r} - \frac{3\al}{\sqrt 2}
    \bm C_z
    + \frac{3\al^2}{2r}z \hat{\bm z}
    }{2\pi r (1+\al^2)}
    . \label{eq:polarization_concrete}
\end{align}
This vector field is schematically plotted on a unit sphere in 
Fig.~\ref{fig:polarization}.
It can be expanded by the order of $\al$.
The $O(1)$ contribution enters from the spin-orbit coupling $\lambda$, which is proportional to $\hat {\bm r}$.
The contribution of $O(\al^1)$ is a leading-order contribution from $\la G^z\ra$, which is rotating around the $z$ axis.
There is also the small $O(\al^2)$ contribution pointing the $z$ direction.

Based on the Landau theory, we expect the temperature dependence $\al \sim \frac {h}{\lambda} \propto \sqrt{T_c-T}$ near the transition temperature.
Hence the $O(\al^1)$ contribution proportional to $\bm C_z$ in $\bm P_S$ reflects the presence of the primary order parameter $\la \bm G \ra$.
However, the $O(\al)$ contribution vanishes in the total charge $\rho_{\rm tot} = \rho - \bm \nabla \cdot \bm P_S$, 
which behaves as $\propto (T_c-T)$ (not square root).
The primary order parameter would be thus veiled.

\begin{figure}[b]
\begin{center}
\includegraphics[width=70mm]{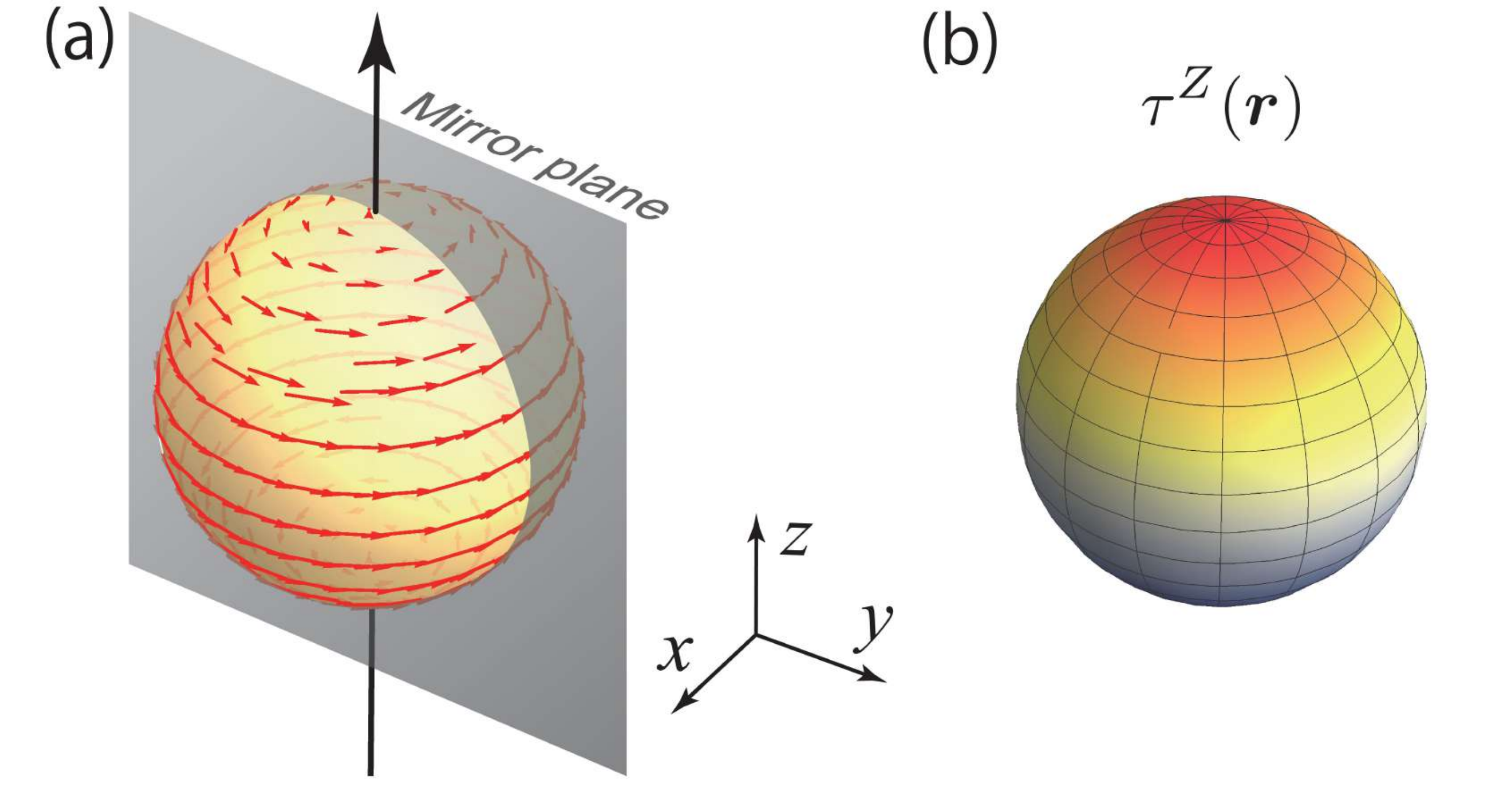}
\caption{
(a) Schematic for the mirror symmetry breaking of $\la \bm P_S(\bm r)\ra$ where the mirror is located parallel to $z$ axis.
(b) Spatial distribution of the chirality density $\la \tau^Z(\bm r)\ra$ on a unit sphere where red (around north pole) and blue (south pole) parts indicate right-chirality-rich and left-chirality-rich regions, respectively.
}
\label{fig:another_map}
\end{center}
\end{figure}

As shown in the $O(\al)$ contribution of Fig.~\ref{fig:polarization}, the $\la G^z\ra$-ordered state breaks the mirror symmetry along the $z$ axis [see Fig.~\ref{fig:another_map}(a)].
The characteristics also appear in the chirality density.
Indeed, its expectation value shows the leading-order $O(\al)$ contribution as follows, 
\begin{align}
    \frac{\la \tau^Z(\bm r) \ra}{R^2(r)} &= \frac{\hbar}{2mc} \,
    \frac{3\sqrt 2 \al z}{2\pi r^2(1+\al^2)} \, .
    \label{eq:tauZ_conc}
\end{align}
Figure~\ref{fig:another_map}(b) shows the spatial distribution of $\la \tau^Z(\bm r) \ra$ on a unit sphere,
which clearly shows the emergence of the chirality dipole distribution.
The red region around the north pole indicates the right-handed electron rich region, while the south pole is the left-handed rich region.
The dipole distribution of $\la \tau^Z(\bm r)\ra$ may also be understood in analogy with the electric dipole distribution.
Namely, the assembly of the local dipoles in $\la \rho\ra$ in solids forms a polar crystal.
Similarly, the assembly of the local chirality-dipole in $\la \tau^Z\ra$ will induce an electric toroidal state in solids characterized by the circulating electric polarization.

Furthermore,
using Eqs.~\eqref{eq:polarization_concrete} and \eqref{eq:tauZ_conc}, 
we arrive at the relation
\begin{align}
    \la \tau^Z\ra &= \frac{4mc}{\imu \hbar^2 e} \, \bm L\cdot \la \bm P_S \ra 
    = - \frac{4mc}{\hbar e} \, \bm r\cdot \big(\bm \nabla \times \la \bm P_S \ra \big) \, ,
        \label{eq:tauz_Ps}
\end{align}
where only the fundamental physical constants appear as the proportional coefficient.
The right-hand side is often used as the definition of the electric toroidal multipoles \cite{Hayami18_JPSJ}.
We note that the right-hand side of Eq.~\eqref{eq:tauz_Ps} directly includes $\bm r$ and hence depends on the choice of the origin, while the chirality in the left-hand side does not.
Thus the $\bm G=\bm \ell \times \bm s$ order parameter is closely related to the microscopic chirality operator $\tau^Z$, which derives from the fundamental Dirac theory and is relevant to the electric toroidal moment.
Chirality is an indispensable degree of freedom to unambiguously describe electronic states.

{\it Summary and discussion.---}
We have investigated the electronic degrees of freedom in the localized electron orbitals using the knowledge of relativistic quantum mechanics.
We clarified that electric toroidal multipoles are microscopically characterized by spin-derived electric polarization and chirality in the Dirac theory, which are closely related to the spin current tensor. 
In particular, the chirality intrinsic to elementary particles is the essence of electric toroidal multipoles.

When considering the ordered state of such electric toroidal multipoles, the corresponding components of the spin current tensor are modulating on an atomic scale.
It would be hard to detect the primary order parameter by conventional spectroscopies.
In this context, it is interesting to study some exotic phase transitions, such as URu$_2$Si$_2$ \cite{Mydosh11} and CeCoSi \cite{Tanida18,Matsumura22}.

Thus, the spin-derived electric polarization and the chirality are fundamental quantities characterizing the quantum states of materials. 
Recently, the importance of electric toroidal monopole in chiral crystals and molecules has been noted \cite{Oiwa22,Kishine22}. 
Mapping out the spin-derived electric polarization and the chirality distributions in these materials is an interesting future challenge.

\section*{Acknowledgement}
We are grateful to H. Kusunose and T. Miki for useful comments. This work was supported by KAKENHI Grants 
No.~19H01842, 
No.~19H05825, 
No.~21H01789, 
No.~21H04437, 
No.~21K03459. 
A part of the numerical calculation was performed in MASAMUNE-IMR of the Center for Computational Materials Science, Institute for Materials
Research, Tohoku University.

\clearpage

\makeatletter
\renewcommand{\thepage}{S\arabic{page}}
\renewcommand{\thesection}{S\arabic{section}}
\renewcommand{\theequation}{S\arabic{equation}}
\renewcommand{\thefigure}{S\arabic{figure}}
\renewcommand{\thetable}{S\arabic{table}}
\makeatother

\setcounter{page}{1}
\setcounter{section}{0}
\setcounter{equation}{0}
\setcounter{table}{0}
\setcounter{figure}{0}

\noindent
{\bf SUPPLEMENTARY MATERIAL FOR \\
``Spin current and chirality degrees of freedom inherent in localized electron orbitals''
}
\\[2mm]
S. Hoshino,
M.-T. Suzuki, and
H. Ikeda
\\[2mm]
(Dated: \today)

\section{Complete multipole basis set}

\subsection{Definition}

We consider the electrons bound on the atom which has the angular momentum $\ell$ ($=1,2,3$) and spin $s=1/2$.
With the atomic spin-orbit coupling, it is appropriate to work with the total angular momentum $j$.
The angular momentum coupling to obtain $j$ is symbolically given by
\begin{align}
    \ell \times \tfrac 1 2 = (\ell+\tfrac 1 2) + (\ell-\tfrac 1 2)
    = \sum_{j=\ell-1/2}^{\ell+1/2} j
    ,
\end{align}
which provides a classification scheme in a one-body Hilbert space with the dimension $(2\ell+1)(2s+1)$, as described by the single-electron creation operator $c^\dg_{m\sg}$ ($m\in [-\ell,\ell]$, $\sg=\pm 1/2$).
The physical quantity is represented by the combination of the creation and annihilation operators ($c_{m\sg}^\dg c_{m'\sg'}$) which have $[(2\ell+1)(2s+1)]^2$ components in total
(i.e., Liouville or extended Hilbert space).
It can also be classified by employing the following symbolic coupling rule:
\begin{align}
&(\ell \times \tfrac 1 2) \times (\ell \times \tfrac 1 2)
\\
&= [(\ell+\tfrac 1 2) + (\ell-\tfrac 1 2)] \times [(\ell+\tfrac 1 2) + (\ell-\tfrac 1 2)]
\\
&= (\ell+\tfrac 1 2)\times (\ell+\tfrac 1 2)
\ \ \ ({\rm Type\ }a)
\nonumber \\
&\ \ 
+ (\ell-\tfrac 1 2)\times (\ell-\tfrac 1 2)
\ \ \ ({\rm Type\ }b)
\nonumber \\
&\ \ 
+ \frac{(\ell+\tfrac 1 2)\times (\ell-\tfrac 1 2) + (\ell-\tfrac 1 2)\times (\ell+\tfrac 1 2)}{2}
\ \ \ ({\rm Type\ }c)
\nonumber \\
&\ \ 
+ \frac{(\ell+\tfrac 1 2)\times (\ell-\tfrac 1 2) - (\ell-\tfrac 1 2)\times (\ell+\tfrac 1 2)}{2\imu}
\ \ \ ({\rm Type\ }d)
\\
&= 
\sum_{p=0}^{2\ell+1} p_a
+ \sum_{p=0}^{2\ell-1} p_b
+ \sum_{p=1}^{2\ell} (p_c + p_d)
,
\end{align}
where the subscripts $a$ and $b$ respectively represent $(\ell+\tfrac 1 2)$- and $(\ell-\tfrac 1 2)$-diagonal components.
Another subscripts $c$ and $d$ indicate the off-diagonal components.

In terms of matrix representation \cite{Chikano21}, 
we first consider a rank-$p$ multipole
\begin{align}
    \mathcal O(p) = 
    \begin{pmatrix}
    O_{11} & O_{12} \\
    O_{21} & O_{22}
    \end{pmatrix}
    ,
\end{align}
where the first (1) and second (2) indices respectively represent $j=\ell+1/2$ and $j=\ell-1/2$ blocks.
There are the Hermiticity relations $O_{11}=O_{11}^\dg$, $O_{22}=O_{22}^\dg$ $O_{12} = O_{21}^\dg$.
The four kinds of $a,b,c,d$ multipoles are then written as
\begin{align}
    \mathcal O(p_a) &= \begin{pmatrix}
    O_{11} &  \\
     & 0
    \end{pmatrix}
    , \\
    \mathcal O(p_b) &= \begin{pmatrix}
    0 & \\
     & O_{22}
    \end{pmatrix}
    , \\
    \mathcal O(p_c) &= \begin{pmatrix}
     & O_{12} \\
    O_{21} & 
    \end{pmatrix}
    , \\
    \mathcal O(p_d) &= \imu \begin{pmatrix}
     & - O_{12} \\
    O_{21} & 
    \end{pmatrix}
    .
\end{align}
For reference, the multipoles for the $\ell=1$ case are listed in the left column of Tab.~\ref{tab:multipolefunction}.
$\mathcal O(p_c)$ and $\mathcal O(p_d)$ are also referred to as plus $(+)$ and minus $(-)$ components of the $j$-off-diagonal contribution.
The concrete forms of the multipoles and the detailed method for their construction are given in Ref.~\cite{Chikano21}.

Our constructed matrix representation for the multipoles have the Hermiticity
\begin{align}
      \mathcal O^{\xi*}_{12} =  \mathcal O^{\xi}_{21}
    , \label{eq:multi_herm}
\end{align}
the orthonormality
\begin{align}
    \sum_{12} \mathcal O^{\xi *}_{12} \mathcal O_{12}^{\xi'}
    = \delta_{\xi\xi'}
    , \label{eq:multi_orth}
\end{align}
and the completeness
\begin{align}
    \sum_\xi \mathcal O^{\xi*}_{12} \mathcal O^{\xi}_{34} = 
    \delta_{13} \delta_{24}
    , \label{eq:multi_comp}
\end{align}
where we have used the short-hand notations as $1 = (m_1\sg_1)$ and $\xi = (p\gm\eta)$.
With these matrices, the multipole operators are defined by
\begin{align}
    X^\gm(p_\eta) &= \sum_{mm'\sg\sg'} 
    [\mathcal O^\gm(p_\eta)]_{m\sg,m'\sg'} 
    c_{m\sg}^\dg  c_{m'\sg'}
    .
\end{align}
We also have the inverse relation
\begin{align}
    c_{m\sg}^\dg c_{m'\sg'} &= \sum_{p\eta \gm} 
    [\mathcal O^\gm(p_\eta)]_{m'\sg',m\sg}
    X^\gm(p_\eta)
    ,
\end{align}
which are used in the main text.

\begin{table*}
    \caption{
    Angular distribution of the multipole basis function $f(\bm r; p_\eta, \gm)$ defined in Eq.~\eqref{eq:def_multipole_basis} for  
    $f=\rho,M_{S\mu},j_{S\mu},P_{S\mu},\tau^{Z},\tau^Y$, and $\bm \nabla \cdot \bm P_S$.
    For the demonstration, we have taken $\ell=1$.
    The blank space indicate zero value.
    }
    \centering
\resizebox{0.96\textwidth}{!}{
    \begin{tabular}{ccc|c|ccc|ccc|ccc|c|c|c}
    \hline
    &&&$\rho$ ($\sim \tau^X$)&$M_{Sx}$&$M_{Sy}$&$M_{Sz}$&$j_x$&$j_y$&$j_z$&$P_{Sx}$&$P_{Sy}$&$P_{Sz}$&$\tau^Z$&$\tau^Y\ (\sim \bm \nabla \cdot \bm M_S)$ & $\bm \nabla \cdot \bm P_S$  \\
    && & $\psi^\dg\psi$ 
    & \multicolumn{3}{c|}{$\psi^\dg \bm \sg \psi$} 
    & \multicolumn{3}{c|}{$\psi^\dg \dvec{\bm p} \psi$} 
    & \multicolumn{3}{c|}{$\psi^\dg \dvec{\bm p}\times \bm \sg \psi$} 
    & $\psi^\dg \dvec{\bm p}\cdot \bm \sg \psi$
    & $\bm \nabla \cdot (\psi^\dg \bm \sg \psi)$
    \\
    \hline
    \multicolumn{2}{c}{Multipoles}& SI/TR & $+/+$
    & \multicolumn{3}{c|}{$ +/-$} 
    & \multicolumn{3}{c|}{$ -/-$} 
    & \multicolumn{3}{c|}{$ -/+$} 
    & $ -/+$
    & $ -/-$
    & $ +/+$
    \\
\hline
&$\mathcal O(0_a)$&$+/+$&$1$&&&&&&&$-x$&$-y$&$-z$&&&$-1$
\\
&$\mathcal O^{x}(1_a)$&$+/-$&&$2r^2-x^2$&$-xy$&$-xz$&&$-z$&$y$&&&&&$x$&
\\
&$\mathcal O^{y}(1_a)$&$+/-$&&$-yx$&$2r^2-y^2$&$-yz$&$z$&&$-x$&&&&&$y$&
\\
&$\mathcal O^{z}(1_a)$&$+/-$&&$-zx$&$-zy$&$2r^2-z^2$&$-y$&$x$&&&&&&$z$&
\\
&$\mathcal O^{xy}(2_a)$&$+/+$&$-xy$&&&&&&&$y$&$x$&&&&$xy$
\\
&$\mathcal O^{yz}(2_a)$&$+/+$&$-yz$&&&&&&&&$z$&$y$&&&$yz$
\\
&$\mathcal O^{zx}(2_a)$&$+/+$&$-zx$&&&&&&&$z$&&$x$&&&$zx$
\\
Diag.
&$\mathcal O^{x^2-y^2}(2_a)$&$+/+$&$-(x^2-y^2)$&&&&&&&$x$&$-y$&&&& $x^2-y^2$
\\
($j=3/2$)
&$\mathcal O^{3z^2-r^2}(2_a)$&$+/+$&$-(3z^2-r^2)$&&&&&&&$-x$&$-y$&$2z$&&&$3z^2-r^2$
\\
&$\mathcal O^{xyz}(3_a)$&$+/-$&&$-yz$&$-zx$&$-xy$&&&&&&&&$-xyz$&
\\
&$\mathcal O^{x(5x^2-3r^2)}(3_a)$&$+/-$&&$r^2-3x^2$&$xy$&$xz$&&&&&&&&$-x(5x^2-3r^2)$&
\\
&$\mathcal O^{y(5y^2-3r^2)}(3_a)$&$+/-$&&$yz$&$r^2-3y^2$&$yz$&&&&&&&&$-y(5y^2-3r^2)$&
\\
&$\mathcal O^{z(5z^2-3r^2)}(3_a)$&$+/-$&&$zx$&$zy$&$r^2-3z^2$&&&&&&&&$-z(5z^2-3r^2)$&
\\
&$\mathcal O^{z(x^2-y^2)}(3_a)$&$+/-$&&$-zx$&$zy$&$y^2-x^2$&&&&&&&&$-z(x^2-y^2)$&
\\
&$\mathcal O^{x(y^2-z^2)}(3_a)$&$+/-$&&$z^2-y^2$&$-xy$&$xz$&&&&&&&&$-x(y^2-z^2)$&
\\
&$\mathcal O^{y(z^2-x^2)}(3_a)$&$+/-$&&$yx$&$x^2-z^2$&$-yz$&&&&&&&&$-y(z^2-x^2)$&
\\ \hline
&$\mathcal O(0_b)$&$+/+$&$1$&&&&&&&$x$&$y$&$z$&&&$1$
\\
Diag.
&$\mathcal O^{x}(1_b)$&$+/-$&&$2x^2-r^2$&$2xy$&$2xz$&&$-z$&$y$&&&&&$x$&
\\
($j=1/2$)
&$\mathcal O^{y}(1_b)$&$+/-$&&$2yx$&$2y^2-r^2$&$2yz$&$z$&&$-x$&&&&&$y$&
\\
&$\mathcal O^{z}(1_b)$&$+/-$&&$2zx$&$2zy$&$2z^2-r^2$&$-y$&$x$&&&&&&$z$&
\\ \hline
&$\mathcal O^{x}(1_c)$&$+/-$&&$x^2-2r^2$&$xy$&$xz$&&$z$&$-y$&&&&&$x$&
\\
&$\mathcal O^{y}(1_c)$&$+/-$&&$yx$&$y^2-2r^2$&$yz$&$-z$&&$x$&&&&&$y$&
\\
&$\mathcal O^{z}(1_c)$&$+/-$&&$zx$&$zy$&$z^2-2r^2$&$y$&$-x$&&&&&&$z$&
\\
Off-diag.
&$\mathcal O^{xy}(2_c)$&$+/+$&$xy$&&&&&&&$y$&$x$&&&&$xy$
\\
$(+)$
&$\mathcal O^{yz}(2_c)$&$+/+$&$yz$&&&&&&&&$z$&$y$&&&$yz$
\\
&$\mathcal O^{zx}(2_c)$&$+/+$&$zx$&&&&&&&$z$&&$x$&&&$zx$
\\
&$\mathcal O^{x^2-y^2}(2_c)$&$+/+$&$x^2-y^2$&&&&&&&$x$&$-y$&&&&$x^2-y^2$
\\
&$\mathcal O^{3z^2-r^2}(2_c)$&$+/+$&$3z^2-r^2$&&&&&&&$-x$&$-y$&$2z$&&&$3z^2-r^2$
\\ \hline
&$\mathcal O^{x}(1_d)$&$+/+$&&&&&&&&&$-z$&$y$&$x$&&
\\
&$\mathcal O^{y}(1_d)$&$+/+$&&&&&&&&$z$&&$-x$&$y$&&
\\
&$\mathcal O^{z}(1_d)$&$+/+$&&&&&&&&$-y$&$x$&&$z$&&
\\
Off-diag.
&$\mathcal O^{xy}(2_d)$&$+/-$&&$zx$&$-zy$&$y^2-x^2$&&&&&&&&&
\\
$(-)$
&$\mathcal O^{yz}(2_d)$&$+/-$&&$z^2-y^2$&$xy$&$-xz$&&&&&&&&&
\\
&$\mathcal O^{zx}(2_d)$&$+/-$&&$-yx$&$x^2-z^2$&$yz$&&&&&&&&&
\\
&$\mathcal O^{x^2-y^2}(2_d)$&$+/-$&&$-zy$&$-zx$&$xy$&&&&&&&&&
\\
&$\mathcal O^{3z^2-r^2}(2_d)$&$+/-$&&$-zy$&$zx$&&&&&&&&&&
\\
\hline
    \end{tabular}
}
    \label{tab:multipolefunction}
\end{table*}

\subsection{Spatial distribution for $\ell=1$}

Table \ref{tab:multipolefunction} summarizes the angular distribution of the multipole basis functions defined in Eq.~\eqref{eq:def_multipole_basis} of the main text for $\ell=1$ as the demonstration.
We write its expression here again:
\begin{align}
    f(\bm r) &= \sum_{p\gm\eta} X^\gm(p_\eta) 
    f(\bm r; p_\eta, \gm)
    ,
\end{align}
where $f$ takes $\rho, \bm M_S, \bm j, \bm P_S, \tau^{Z},\tau^Y$, and $\bm \nabla\cdot \bm P_S$.

We point out that the current density $\bm j$ vanishes for the multipole $X^\gm(3_a)$.
This is because the highest rank $3$ ($=2\ell+1$) is made only when the coupling to spin is considered.
Hence, the current, which does not have apparent contribution from the spin, cannot have the finite value for the highest-rank magnetic multipole.

\section{Relativistic quantum mechanics}

\subsection{Non-relativistic limit}

Here we summarize the Hamiltonian for the electrons in solids in the non-relativistic limit.
We consider the Hamiltonian in the second quantization form up to the order of $c^{-2}$ \cite{Berestetskii_book}:
\begin{align}
    \mathscr H &= 
    \int \diff \bm r \, \psi^\dg \Big[
    \frac{1}{2m} \qty( \bm p - \frac e c \bm A )^2
    + e\Phi
    -\frac{\bm p^4}{8m^3 c^2}
    \nonumber \\
    &{\ \ \ }    
- \frac{\hbar e}{2mc} \bm B \cdot \bm \sg
    - \frac{\hbar e}{4m^2c^2} \bm \sg \cdot (\bm E \times \bm p)
- \frac{\hbar^2 e}{8m^2c^2} {\rm div\,} \bm E
    \Big] \psi
    , \label{eq:Pauli_Hamiltonian}
\end{align}
where $\bm p = -\imu \hbar \bm \nabla$ is the momentum operator and
$\displaystyle \psi = \begin{pmatrix} \psi_\ua \\ \psi_\da\end{pmatrix}$ is the annihilation operator of electrons.
For example, if we adopt the component representation, the spin operator can be written as $\psi^\dg \sg^\mu \psi = \sum_{\sg\sg'}\psi^\dg_\sg \sg^\mu_{\sg\sg'} \psi_{\sg'}$.
The magnetic and electric fields are $\bm B = \bm \nabla \times \bm A$ and $\bm E = -\bm \nabla\Phi$, respectively.

The electric charge and current are given by the functional derivative of the Hamiltonian: 
$\bm j_{\rm tot} = -\frac{\delta \mathscr H}{\delta (\bm A/c) } $
and
$\rho_{\rm tot} = \frac{\delta \mathscr H}{\delta \Phi } $.
The explicit forms are given by
\begin{align}
    \bm j_{\rm tot} &= 
    - \frac{\imu \hbar e}{2m} \psi^\dg  \dvec{\bm \nabla} \psi
    - \frac{e^2}{mc} \psi^\dg \psi \bm A
    + \frac{\hbar e}{2m} \bm \nabla \times (\psi^\dg \bm \sg \psi)
    \label{eq:current_exp}
\end{align}
for the current density, and
\begin{align}
    \rho_{\rm tot} &= 
    e \psi^\dg \psi 
    + \frac{\hbar^2 e}{8m^2c^2} \Big[ \imu \bm \nabla \cdot (\psi^\dg  \dvec{\bm \nabla} \times \bm \sg \psi)
    + \bm \nabla^2 (\psi^\dg \psi)   \Big]
    \label{eq:charge_exp}
\end{align}
for the charge density.
These quantities serve as a source term in the Maxwell equation.

\begin{figure}[t]
\begin{center}
\includegraphics[width=75mm]{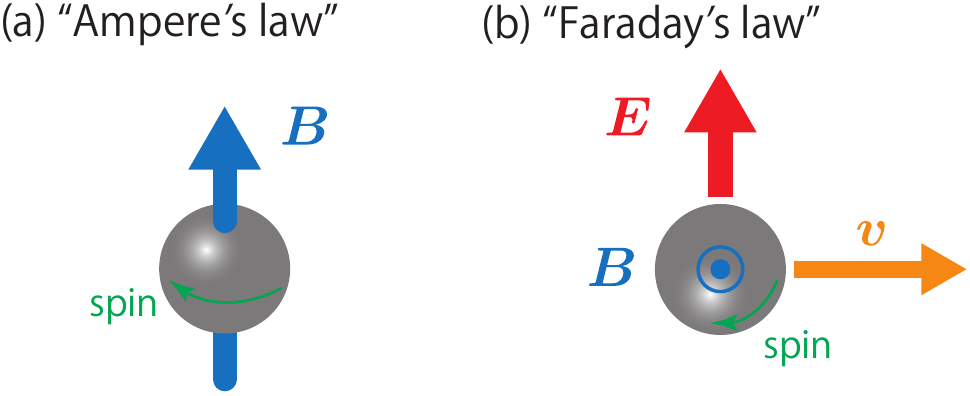}
\caption{
Cartoon illustrations for (a) the microscopic magnetization $\bm M_S$ and (b) the microscopic electric polarization $\bm P_S$, where the electron is schematically drawn as a gray sphere which is negatively charged and has a radius of the order of the Compton length $\sim \hbar / mc$. 
}
\label{fig:faraday}
\end{center}
\end{figure}

With these expressions we obtain $\rho$, $\bm j$, $\bm M_S$ and $\bm P_S$ defined in the main text.
We can intuitively understand the physical origin of these quantities by considering the motion of the charged sphere.
As shown in Fig.~\ref{fig:faraday}(a), the spinning motion ($\bm \sg$) of the charged sphere generates a magnetic field by Ampere's law to produce a magnetization.
As for the electric polarization $\bm P_S$, its expression in Eq.~\eqref{eq:p_def} is interpreted as $\bm v \times \bm \sg$ with $\bm v$ being a velocity of the sphere.
A drifting spinning sphere shown in Fig.~\ref{fig:faraday}(b) produces the electric field by Faraday's law to generate electric polarization.
One may also recognize that this situation is analogous to the vortex dynamics in superconductors which is responsible for the production of resistivity \cite{Tinkham_book}.

\subsection{Connection to the Dirac equation}
It is useful to make an explicit connection to the Dirac theory.
The Schr\"{o}dinger equation is given by \cite{Berestetskii_book}
\begin{align}
    \imu \hbar \frac{\partial \Psi}{\partial t} &=  \qty[
    c \bm \al \cdot \qty( \bm p- \frac e c \bm A)
    + \beta mc^2 
    + e\Phi
    ] \Psi
    ,
\end{align}
where $\Psi = 
\begin{pmatrix}
\phi \\ \chi
\end{pmatrix}
$ is the Dirac's four-component spinor.
The $4\times 4$ matrices are defined by
\begin{align}
    \bm \al = 
    \begin{pmatrix}
    0 & \bm \sg \\
    \bm \sg & 0
    \end{pmatrix}
    ,\ \ \ 
    \beta = 
    \begin{pmatrix}
    1 & 0 \\
    0 & -1
    \end{pmatrix}
    .
\end{align}
For convenience we also define the gamma matrices $\gm^0 = \beta$, $\gm^1 = \beta \al^x$, $\gm^2 = \beta \al^y$, $\gm^3 = \beta \al^z$ ($\bm \gm = \beta \bm \al$).
The electric current and charge densities are given by
\begin{align}
    \bm j &= 
    e c  \Psi^\dg \bm \alpha \Psi
    = ec \bar \Psi \bm \gm \Psi
    , \\
    \rho &= 
    e \Psi^\dg \Psi
    = e\bar \Psi \gm^0 \Psi
    ,
\end{align}
where $\bar \Psi = \Psi^\dg \gm^0$.
Taking the non-relativistic limit and employing the second-quantized form, one can obtain the expressions in the last subsection.
For the derivation, the norm conservation needs to be considered.

As discussed in the main text, another important physical quantity related to the electric polarization is the chirality operator
\begin{align}
    A^0 &= \bar \Psi\gm^0\gm^5\Psi= \Psi^\dg \gm^5 \Psi
    , \\
    \gm^5 &= -\imu \gm^0 \gm^1 \gm^2 \gm^3 =
    \begin{pmatrix}
    0 & -1 \\
    -1 & 0
    \end{pmatrix}
    .
\end{align}
Defining the right- and left-handed Weyl particles by $\phi_R = (\phi+\chi)/\sqrt 2$ and $\phi_L = (\phi-\chi)/\sqrt 2$, we have another expression:
\begin{align}
 A^0 = |\phi_L|^2 - |\phi_R|^2   , 
\end{align}
which clearly measures the chirality of the electrons at each position.

More generally, the bilinear forms made from the Dirac spinor have $4\times4=16$ components \cite{Berestetskii_book}.
For condensed matter physics, it is useful to summarize the non-relativistic limit of these quantities.
The full list is given as follows:
\begin{align}
    S = \bar \Psi\Psi &\simeq |\phi|^2
    \\
    P = \imu \bar \Psi\gamma^5\Psi &\simeq -\frac{\hbar}{2mc} \bm \nabla \cdot (\phi^*\bm \sg\phi)
    \\
    V^0 = \bar \Psi\gm^0\Psi &\simeq |\phi|^2 
    \\
    \bm V = \bar \Psi\bm \gm\Psi &\simeq \frac{\hbar}{2mc}
    \big[-\imu \phi^*\dvec{\bm \nabla} \phi + \bm \nabla \times (\phi^* \bm \sg \phi)\big]
    \\
    A^0 = \bar \Psi\gm^0\gm^5\Psi &\simeq \frac{\imu\hbar}{2mc}\phi^*\dvec{\bm \nabla}\cdot \bm \sg \phi
    \\
    \bm A = \bar \Psi\bm \gm\gm^5\Psi &\simeq  -\phi^*\bm \sg \phi
    \\
    \bm T_1 = \imu \bar \Psi\bm \al\Psi &\simeq \frac{\hbar }{2mc} (\imu\phi^*\dvec{\bm \nabla}\times \bm \sg \phi + \bm \nabla |\phi|^2)
    \\
    \bm T_2 = 
    \bar \Psi \bm \al \gm^5 \Psi
    &\simeq -\phi^*\bm \sg \phi
\end{align}
The symbols in the most left side originate from a Lorentz scalar ($S$), pseudoscalar ($P$), four-compoennt current vector ($V$), pseudovector (axial current, $A$), and antisymmetric tensor ($T$), respectively.
In the most right-hand sides,
the electromagnetic field is neglected and only the leading-order contributions in $1/c$ expansion are kept.
There is a rough correspondence between the above bilinear forms and quantities defined in the main text: 
$V^0 \leftrightarrow \rho$, 
$\bm V \leftrightarrow \bm j$,
$\bm T_2 \leftrightarrow \bm M_S$,
$\bm T_1 \leftrightarrow \bm P_S$,
$A^0 \leftrightarrow \tau^Z$,
$S \leftrightarrow \tau^X$,
$P \leftrightarrow \tau^Y$.
Note that the second quantization form is used in the main text.

\vspace{10mm}
\noindent
{\bf \large References}
\\[1mm]
See the list of references in the main text.


\begin{thebibliography}{99}
\bibitem{matrixO} 
The matrix representations of the inter-$j$ components are given by
$\mathcal O(p_c)=\begin{pmatrix} & O_{12} \\ O_{12}^\dagger & \end{pmatrix}$, 
$\mathcal O(p_d)=\imu \begin{pmatrix} & -O_{12} \\ O_{12}^{\dagger} & \end{pmatrix}$.
Here, $O_{12}$ represents an off-diagonal block connecting different $j=\ell\pm 1/2$ multiplets.
See Ref.~\cite{suppl} for more details.
\end{thebibliography}
\end{document}